\documentclass[aip,preprint,superscriptaddress,amsmath,amssymb,pof]{revtex4-1}
\usepackage{graphicx} 
\usepackage{dcolumn}
\usepackage{bm}
\usepackage{color}

\begin{document}

\title{Variable energy flux in quasi-static magnetohydrodynamic turbulence}

\author{Mahendra K. Verma}
\affiliation{Department of Physics, Indian Institute of Technology -- Kanpur 208016, India}
\email[Email : ]{mkv@iitk.ac.in}

\date{\today}

\begin{abstract}
Experiments and numerical simulations show that the energy spectrum of the magnetohydrodynamic turbulence in the quasi-static limit deviates from Kolmogorov's  $k^{-5/3}$ spectrum as the external magnetic field, or equivalently the interaction parameter, is increased.  To explain this phenomena, we  construct an analytical turbulence model with variable energy flux that arises due to the Lorentz-force induced dissipation.   The energy spectra computed using our model for various interaction parameters are in qualitative agreement with earlier experimental and numerical results.  
\end{abstract}

%\pacs{91.25.Cw, 52.65.Kj, 47.20.Ky}
%91.25.Cw---Origins and models of the magnetic field; dynamo theories 
%52.65.Kj---Magnetohydrodynamics and fluid equation 
%47.20.Ky---Nonlinearity, bifurcation, and symmetry breaking 

\maketitle

\section{Introduction}
Magnetohydrodynamic (MHD) flows are observed in two kinds of fluids: plasma and liquid metals.  The flows in the stars and galaxies involve MHD plasma, while the interiors of the planet including the Earth, fission and fusion reactors (e.g., International Thermonuclear Experimental Reactor), industrial processes like metal plate rolling and crystallization etc. have liquid metals as the operational fluid.    An important characteristic of the liquid metals  is its low magnetic Prandtl number $\mathrm{Pm}$, which is the ratio of the kinematic viscosity $\nu$ and the magnetic diffusivity $\eta$.  Consequently, in many situations, especially in industrial applications, the magnetic  Reynolds number $\mathrm{Rm} = U L/\eta$  ($U,L$ are the large scale velocity and length scales respectively)  is quite small.  Therefore, liquid metal flows under an idealized limit $\mathrm{Rm} \rightarrow 0$, which is referred to as ``quasi-static limit",  is of major interest to scientists and technologists.

In the quasi-static limit, the MHD equations get simplified significantly.  Here, the induced magnetic field tends to be very small  because of the large magnetic diffusivity, and it gets slaved to the velocity field that yields the Lorentz force as
\begin{equation}
{\bf F} = -\frac{\sigma B_0^2}{\rho} \frac{1}{\nabla^2} \frac{\partial^2 {\bf u}}{\partial z^2},
\end{equation}
where $\rho$ is the density of the fluid, ${\bf u}$ is the velocity field, and ${\bf B} = B_0 \hat{z}$ is the external uniform magnetic field~\cite{Davidson:MHD:book,Moreau:book}.   The Lorentz force induces additional dissipation of the kinetic energy.   We will show in the present paper that the dissipation due to the Lorentz force reduces the inertial-range energy flux, and makes it scale-dependent.   As a result, the energy spectrum gets steepened as the interaction parameter, which is the ratio of the Lorentz force and the nonlinear term ${\rho \bf u \cdot \nabla u}$, is increased.

Several experimental and numerical simulations have been performed to study quasi-static MHD turbulence under a strong external magnetic field (see Knaepen and Moreau~\cite{Knaepen:2008ARFM} and references therein).  Kolesnikov and Tsinober~\cite{Kolesnikov:1974} and Alemany {\it et al.}~\cite{Alemany:1979JM} performed experiments on mercury for low $\mathrm{Rm}$. They reported that the energy spectrum for the velocity field is $k^{-3}$ when the interaction parameter is significant.  Kit and Tsinoner~\cite{Kit:1971} argued that the observed spectrum is due to the  two-dimensional nature of the flow, while Alemany {\it et al.}~\cite{Alemany:1979JM} obtained $k^{-3}$ spectrum by equating the nonlinear time scale with the Joule time scale, which was assumed to be wavenumber-independent.  Sommeria and Moreau~\cite{Sommeria:1982JFM} provided phenomenological arguments for identifying the time scales when the  MHD turbulence with low magnetic Reynolds number becomes two-dimensional.

Numerical simulations have been applied to study liquid metal flows under the quasi-static approximation or for low magnetic Reynolds number.  Many numerical simulations~\cite{Zikanov:1998JFM,Vorobev:2005PF,Ishida:2007PF,Knaepen:2008ARFM,Favier:2010PFb,Favier:2011JFM} show steepening of the energy spectrum for large interaction parameter, similar to those seen in the experiments.    It has been observed that for large interaction parameters, the flow becomes  anisotropic with energy concentrated near the plane perpendicular to the external magnetic field~\cite{Zikanov:1998JFM,Ishida:2007PF,Burattini:2008PF,Favier:2010PFb,Favier:2011JFM}.   Ishida and Kaneda~\cite{Ishida:2007PF} studied studied the anisotropic velocity correlation spectrum using theoretical and numerical tools.   Favier {\it et al.}~\cite{Favier:2010PFb} argued that the flow is not entirely two-dimensional due to the presence of all the three components of the velocity field near the equator.  This kind of fluid configuration is called ``two-and-a-half-dimensional" or ``two-dimensional and three-component" turbulence.  Favier {\em et al.}~\cite{Favier:2011JFM} also applied eddy-damped quasi-normal Markovian (EDQNM) model to study anisotropic turbulence in the quasi-static limit, and observed consistency with the numerical results.  Burattini {\it et al.}~\cite{Burattini:2008PD} showed that the nonlinear energy transfer at small scales is both radial and angular, thus necessitates an anisotropic treatment of the the quasi-static MHD turbulence. 

In this paper we derive an expression for the energy flux that includes the dissipation induced by the Lorentz force.  We show that the energy flux has a strong wavenumber dependence that leads to the steepening of the energy spectrum.    It is important to emphasize that  we use isotropic Kolmogorov-like energy spectrum in the present model.  The angular dependence of the spectrum would be incorporated in future.

The organization of the paper is as follows.  We describe the model and its results in Sections 2 and 3 respectively.  Section 4 contains conclusions.

\section{The Model}
The equation of motion for the liquid metal flow under the quasi-static approximation is~\cite{Davidson:MHD:book,Moreau:book}
\begin{equation}
\frac{\partial {\bf u}}{\partial t} + {\bf u \cdot \nabla u}  = -\nabla (p/\rho) 
	 -\frac{\sigma B_0^2}{\rho} \frac{1}{\nabla^2} \frac{\partial^2 {\bf u}}{\partial z^2} + \nu \nabla^2 {\bf u},
\end{equation}
along with the incompressibility constraint $\nabla \cdot {\bf u} = 0$ (i.e., $\rho$ is a constant).  In the above equation, ${\bf u}$ is the velocity field, $p$ is the pressure field, ${\bf B}_0$ is the mean magnetic field along the $z$ direction, $\sigma$ is the electrical conductivity of the fluid, and $\nu$ is  the kinematic viscosity of the fluid.    It is convenient to describe the flow properties in the Fourier space, for which we rewrite the Navier-Stokes equation in Fourier space as
\begin{equation}
\frac{\partial u_i ({\bf k)}} {\partial t}   = -i  k_i p({\bf k}) /\rho - i k_j \sum u_j ({\bf q}) u_i( {\bf k-q})
	-\frac{\sigma B_0^2}{\rho} (\cos^2 \theta) u_i - \nu k^2  u_i,
\end{equation}
where $\theta$ is the angle between the mean magnetic field and the wavenumber ${\bf k}$.   The interaction parameter $N$, which is the ratio of the Lorentz force and the nonlinear term, is $\sigma B_0^2 L/(\rho U)$.   It is also important to keep in mind that the Lorentz force in the polar region is stronger than that in the equatorial region due to the $\cos^2\theta$ factor.  Hence, the interaction parameter is an averaged measure. 

The corresponding energy equation is
\begin{equation}
\frac{\partial E ({\bf k})} {\partial t}   = T ({\bf k}) - 2 \frac{\sigma B_0^2}{\rho} \cos^2 (\theta) E ({\bf k}) - 2 \nu k^2  E ({\bf k}),
\end{equation}
where $E ({\bf k}) = |{\bf u(k)}|^2/2$ is the energy spectrum, and $T({\bf k})$ is the kinetic energy transfer rate.   The second and the third terms in the RHS are the dissipation rates due to the Lorentz force and the viscous force respectively.  For $N=0$ and large Reynolds number $\mathrm{Re}=UL/\nu$, the flow becomes turbulent, and the energy spectrum exhibits the famous Kolmogorov's $k^{-5/3}$ power law in the inertial range.  However for finite $N$, the Lorentz force induces additional dissipation that leads to the modification of the energy spectrum.   The Lorentz force is dissipative in the quasistatic limit, so we can define another Reynolds number $\mathrm{Re}_\sigma$ as the ratio of the nonlinear term $\rho {\bf u \cdot \nabla u}$ and the Lorentz force (of dissipative nature), i.e., 
\begin{equation}
\mathrm{Re}_\sigma = \frac{\rho U}{\sigma B_0^2 L} = \frac{1}{N}.
\end{equation}
We term $\mathrm{Re}_\sigma$ as the ``Reynolds number based on resistivity".   We will show below that  $\mathrm{Re}_\sigma$ plays an important role in the dynamics of liquid metals MHD.

\begin{figure}
\begin{center}
\includegraphics[scale=1]{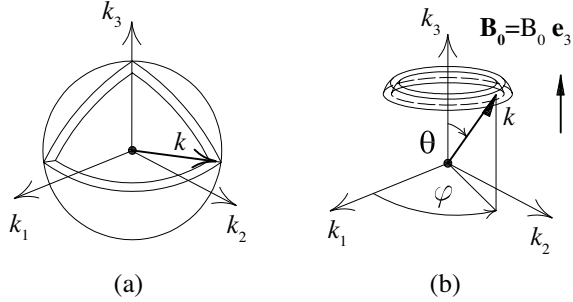}
\caption{(a) Shell decomposition of the wavenumber space; (b) Ring decomposition in which a shell is further divided into rings. Adapted from Teaca {\em et al.}~\cite{Teaca:2009PRE}.}
\label{fig:ring}
\end{center}
\end{figure}
Since the mean magnetic field induces anisotropy~\cite{Zikanov:1998JFM,Burattini:2008PF,Favier:2010PFb}, the local energy density is best described using ring spectrum $E(k,\theta)$~\cite{Teaca:2009PRE,Burattini:2008PD},  where $k$ is the wavenumber of the ring, and $\theta$ is the angle between the mean magnetic field and the wavenumber ${\bf k}$ of the ring, as shown in Fig.~\ref{fig:ring}. We extrapolate Pope's prescription~\cite{Pope:book} for the energy spectrum in isotropic turbulence to the ring spectrum as
\begin{equation}
E(k,\theta) = C (\Pi(k))^{2/3} k^{-5/3}  f_L(k L) f_\eta(k \eta) \frac{g(\theta)}{\pi}
\label{eq:Ek}
\end{equation}
where $C\approx 1.5$ is the Kolmogorov's constant, $\Pi(k)$ is the energy flux emanating  from the wavenumber sphere of radius $k$, and $g(\theta)$ represents the anisotropic component of the energy spectrum. The functions $f_L(kL)$,$f_\eta(k \eta)$ specify the components of the  large-scale and dissipative-scale spectra respectively, and they have been prescribed by Pope~\cite{Pope:book} as
\begin{eqnarray}
f_L(kL) & = & \left( \frac{kL}{[(kL)^2 + c_L]^{1/2}} \right)^{5/3+p_0}, 
\label{eq:fL} \\
f_\eta(k\eta) & = & \exp \left[ -\beta \left\{ [ (k\eta)^4 + c_\eta^4 ]^{1/4}   - c_\eta \right\} \right],
\label{eq:feta}
\end{eqnarray}
where the $c_L, c_\eta, p_0, \beta$ are constants, which are determined by matching the above function with experimental results on isotropic turbulence at high Reynolds number.  We take  $C_L \approx 6.78$, $c_\eta \approx 0.40$, $\beta \approx 5.2$ and $p_0 =2$ as suggested by Pope~\cite{Pope:book}.    We remark that the splitting of $E(k,\theta)$ into a product of  functions of $k$ and $\theta$ is an assumption that needs to be verified by simulations and/or experiments.
The quasi-static MHD turbulence  is anisotropic for moderate to high $N$'s, hence the form of $f_L(kL)$ and $f_\eta(k\eta)$ described by Eqs.~(\ref{eq:fL}) and (\ref{eq:feta}) may not be applicable to quasi-static MHD turbulence.  Yet, in the absence of extensive data sets for such flows, we choose Pope's prescription for our calculations.  In the present paper we focus on the inertial and dissipative range, $f_L(kL) = 1$.   Under the special case of isotropic turbulence, $g(\theta) = 1$, and 
\begin{equation}
\int_0^{\pi}  d\theta E(k,\theta) = E(k) = C (\Pi(k))^{2/3} k^{-5/3}  f_\eta(k \eta)
\end{equation}

In Kolmogorov's phenomenology of fluid turbulence, the energy flux $\Pi(k)$ is a constant in the inertial range since the dissipative term $2\nu k^2 E(k,\theta)$ is negligible.  In quasi-static MHD, the Lorentz force introduces an additional dissipation term.  In contrast to the viscous dissipation, the dissipation due to the Lorentz force is active at all wavenumbers.  As a consequence, the inertial-range energy flux $\Pi(k)$ decreases significantly with the increase of $k$.  Quantitively, the difference between energy fluxes $\Pi(k+dk)$ and $\Pi(k)$ is due to the energy dissipation in the shell $(k,k+dk)$, i.e., 
 \begin{equation}
 \Pi(k+dk) - \Pi(k) = - \left\{ \int_0^\pi d\theta \left[ 2 \nu  k^2 + 2 \frac{\sigma B_0^2}{\rho} \cos^2 \theta \right] E(k,\theta) \right\} dk,
\end{equation}
or
 \begin{equation}
 \frac{d\Pi(k)}{dk} = -\left[ 2c_1  \nu k^2 +2c_2 \frac{\sigma B_0^2}{\rho}  \right] C (\Pi(k))^{2/3} k^{-5/3} f_\eta(k \eta)
 \label{eq:dPidk}
\end{equation}
with 
 \begin{equation}
c_1 =\frac{1}{\pi} \int_0^\pi g(\theta) d \theta; \hspace{0.5cm} 
c_2 = \frac{1}{\pi} \int_0^\pi g(\theta) \cos^2 \theta d \theta.
\end{equation}

We integrate the above equation from $k=k_1$, which is the starting wavenumber of the inertial range.  Assuming that the energy flux at this wavenumber is $\Pi_0$, we obtain
\begin{eqnarray}
\left[ \frac{\Pi(k)}{\Pi_0} \right]^{1/3} & = & 1- \frac{2C c_1}{3} 
			\left( \frac{\nu^3}{\Pi_0 \eta^4} \right)^{1/3} I_1(k \eta)
			-\frac{2 c_2 C \sigma B_0^2}{3\rho} \frac{\eta^{2/3}}{\Pi_0^{1/3}} I_2(k\eta) 
			\nonumber \\
		& = & 1-\frac{2c_1 c_3 C  }{3}   I_1(k \eta) - \frac{2}{3}\frac{c_2 C N}{\sqrt{c_3 Re}}  N  I_2(k\eta),
		\label{eq:flux_exact}
\end{eqnarray}
 where $\eta$ is the Kolmogorov length, and $(\nu^3/\Pi_0 \eta^4)^{1/3} = c_3$, or 
\begin{equation}
\eta =  c_3^{-3/4} \left(\frac{\nu^3}{\Pi_0} \right)^{1/4}.
\end{equation}
We choose $c_3 = 3.1$ in order to achieve $\Pi(k) \rightarrow 0$ for $k \eta \gg 1$ when $N=0$ (isotropic case).  We also take Kolmogorov's constant $C$ to be 1.5 for all our calculations.  The constants $c_1$ and $c_2$ depend quite crucially on $g(\theta)$, which is a function of $N$.  Unfortunately, at present, $g(\theta)$ is not known accurately. Therefore, we choose $c_1=1$ and $c_2=1/2$, which are the values for the isotropic or $N=0$ case ($g(\theta)=1$).   The integrals $I_1$ and $I_2$ in the above equations are
\begin{eqnarray}
I_1(k \eta) & = & \int_{k_1 \eta}^{k \eta} dk' k'^{1/3} f_\eta(k' ) \\
I_2(k \eta) & = & \int_{k_1 \eta}^{k \eta}  dk' k'^{-5/3} f_\eta(k').
\end{eqnarray}
Equation (\ref{eq:flux_exact}) indicates that the second term of the energy flux which arises from the Lorentz force depends crucially on the constant $N'= 2 c_2 C N/(3\sqrt{c_3 Re})$, termed as the ``normalized  interaction parameter".  

Following Pope~\cite{Pope:book}, we assume that the  inertial range wavenumber starts at around $k_1 = 6 \times 2 \pi/L$.   Therefore, the lower limit of the integral is $k_1 \eta = 6 (2 \pi) (\eta/L) = 12 \pi \times  (c_3 Re)^{-3/4} $.   Note that the energy flux $\Pi(k)$ peaks at  $k=k_1$ with value $\Pi_0$. In the limit $\nu \rightarrow 0$, and for  $k$'s in the inertial range, we can approximate the energy flux as
\begin{equation}
\left[ \frac{\Pi(k)}{\Pi_0} \right]^{1/3} \approx 1 - \frac{c_2 C N}{\sqrt{c_3 Re}} \left[ (k_1 \eta)^{-2/3} -(k \eta)^{-2/3}  \right]  \label{eq:approximate_flux}
\end{equation}
which demonstrates that $\Pi(k)$ decrease as $(A-B k^{-2/3})^3$ in the inertial range.  Wavenumber-dependence of the energy flux is consistent with the numerical findings of Ishida and Kaneda~\cite{Ishida:2007PF}.

The energy spectrum can be easily computed from the aforementioned $k$-dependent energy flux using
\begin{equation}
E(k) =
\begin{cases} 
C \Pi_0^{2/3} k^{-5/3}   f_\eta(k \eta) \left[ \frac{\Pi(k)}{\Pi_0} \right]^{2/3}, & \text{if  $k>k_1,$}  \\
C \Pi_0^{2/3} k^{-5/3}  f_L(k L), & \text{otherwise.}
\end{cases}
\end{equation}
Given the energy spectrum, the dissipation rates due to the viscous and Lorentz forces could also be computed as $D_\nu = \int_0^\infty 2\nu k^2 c_1 E(k) dk$ and $D_N = \int_0^\infty (2\sigma B_0^2/\rho) c_2 E(k)dk$ respectively.

\section{Results}
\begin{figure}
\begin{center}
\includegraphics[scale=0.35]{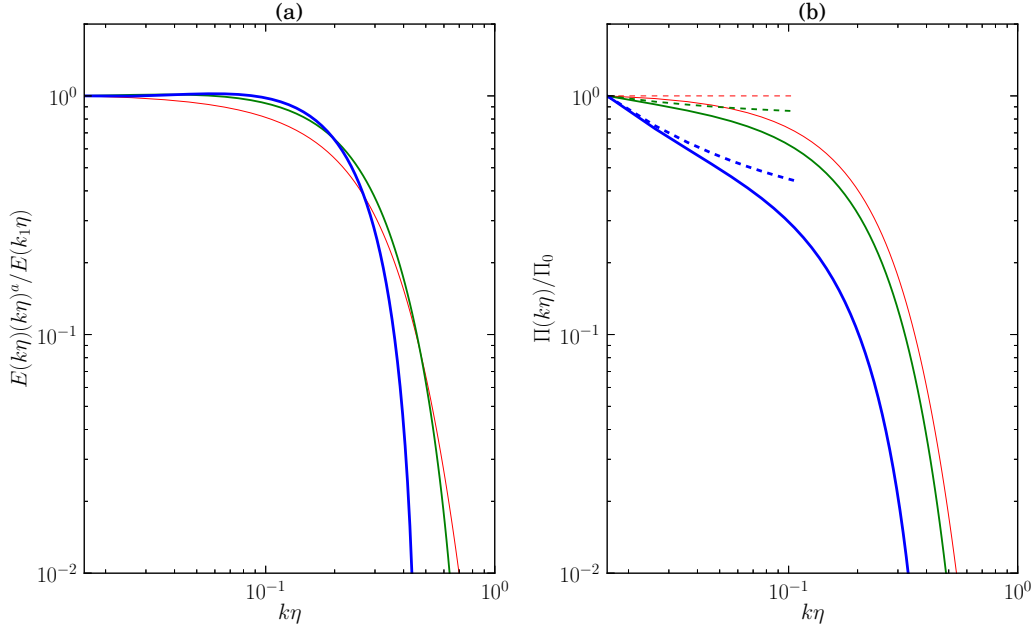}
\caption{  (a) Plots of normalized energy spectra $E(k\eta)(k\eta)^a/E(k_1\eta)$ for $Re=10^4$ and $N=0$ (red thin curve), 1 (green thicker curve), and 5 (thickest blue line).  Here $a=5/3, 1.8$, and 2.1 are the best fit spectral exponents.   (b) Plots of the corresponding energy fluxes (solid lines) and the approximate energy fluxes computed using Eq.~(\ref{eq:approximate_flux}) (dashed lines).} 
\label{fig:Re1e4}
\end{center}
\end{figure}
After the above discussion on the variable energy flux and the energy spectrum, we compute these quantities for two sets of parameters: $Re=10^4$ and $N=0,1,5$; and $Re=2500$ and $N=0,1,5$.  We test whether our model provides results consistent with earlier experiments and numerical simulations.    For $R=10^4$ and $N=0,1,5$, Figure~\ref{fig:Re1e4}(a) exhibits the normalized energy spectra $E(k\eta)(k\eta)^a/E(k_1\eta)$, where $a$ is the spectral exponent.   Figure~\ref{fig:Re1e4}(b) shows the  corresponding normalized energy fluxes $\Pi(k)/\Pi_0$ (solid curves) and the approximate energy fluxes (dashed curves) computed using Eq.~(\ref{eq:approximate_flux}).  These figures show that the energy spectrum follows a power law in the inertial range.  Also, the approximate energy flux computed using Eq.~(\ref{eq:approximate_flux}) matches very well with the actual energy flux computed using Eq.~(\ref{eq:flux_exact}) for  $k\eta$ in the inertial range.  

\begin{figure}
\begin{center}
\includegraphics[scale=0.35]{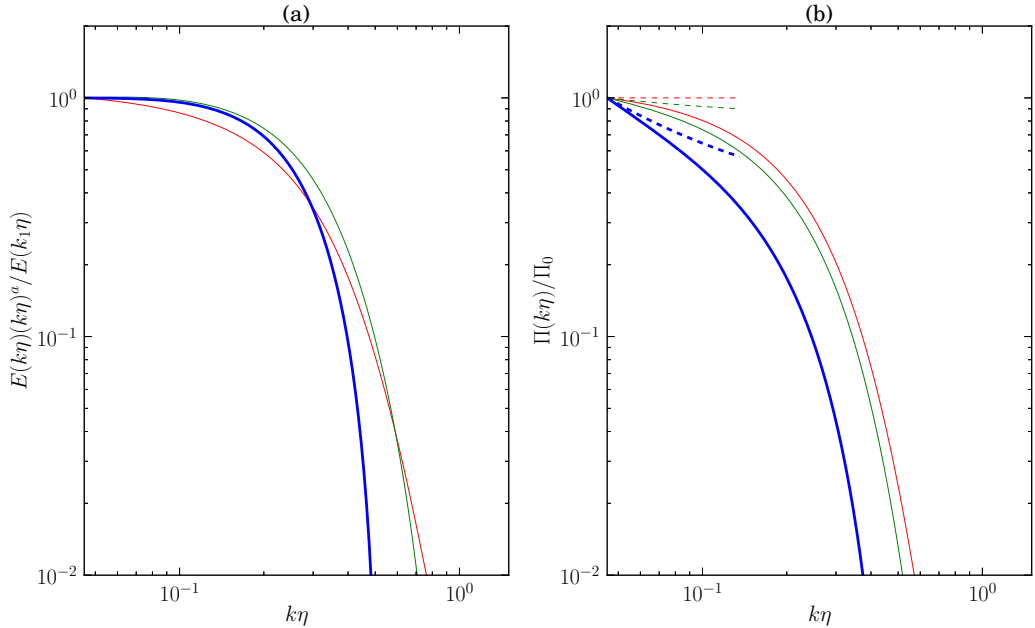}
\caption{  (a) Plots of normalized energy spectra $E(k\eta)(k\eta)^a/E(k_1\eta)$ for $Re=2500$ and $N=0$ (red thin curve), 1 (green thicker curve), and 5 (thickest blue line).  Here $a=5/3, 1.9$, and 2.2 are the best fit spectral exponents.   (b) Plots of the corresponding energy fluxes (solid lines) and the approximate energy fluxes computed using Eq.~(\ref{eq:approximate_flux}) (dashed lines).}
\label{fig:Re2500}
\end{center}
\end{figure}

Figure~\ref{fig:Re2500} exhibits corresponding the energy spectra and fluxes for $Re=2500$ and $N=0,1,5$.  As evident from the normalized energy spectra plots, power law fits reasonably well with the energy spectrum for a relatively narrower inertial range.  The approximate energy flux formula [Eq.~(\ref{eq:approximate_flux})] do not match with the energy flux of Eq.~(\ref{eq:flux_exact}) because the flows with $\mathrm{Re}=2500$ have a relatively smaller inertial range.  Also note that the magnitude of the spectral exponents $a$ for $\mathrm{Re}=2500$ is higher than that for $\mathrm{Re} = 10^4$ because of the increase in the dissipation for lower $\mathrm{Re}$.

%\begin{figure}
%\begin{center}
%\includegraphics[scale=0.30]{figures/exponent.eps}
%\caption{(a) Plots of the spectral index as a function of $N$ for $Re=10^4$ (solid blue curve) and for $Re=1000$ (dashed red curve); (b) Plots of $D_N/D_\nu$ for $Re=10^4$ (solid blue curve) and for $Re=1000$ (dashed red curve) on a log-log scale. }
%\label{fig:avsN}
%\end{center}
%\end{figure}
We compile the spectral indices and the ratio of the dissipation rates $D_N/D_\nu$ for all our runs in Table 1.  The dissipation rate due to the Lorentz force ($D_N$) is significant, in fact, it is more than $D_\nu$ for the parameters studied in the paper.  We should however keep in mind that the system is anisotropic, and the angular dependence is important.  For $N>1$, the dissipative Lorentz force is stronger than the nonlinear term, or $\mathrm{Re_\sigma} < 1$.  Yet, $E(k)$ exhibits power law scaling possibly because the Lorentz force is active at all scale.     The spectral index, as well as the the ratio of the dissipation rates $D_N/D_\nu$, grow monotonically with the interaction parameter $N$.   This trend holds qualitatively in experiments and simulations. 
 \begin{table}[ht]
\caption{Table depicting the spectral exponent $a$ and $D_N/D_\nu$, the ratio of dissipation due to the Lorentz force and the viscous force, for various Reynolds number $\mathrm{Re}$ and interaction parameter $N$.  The normalized interaction parameter $N'= (2/3) (c_2 C N)/\sqrt{c_3 Re}$, where $C$ is the Kolmogrov's constant, $\mathrm{Re}$ is the Reynolds number, and $c_2$ is a parameter.  $\mathrm{Re}_\sigma = 1/N$ is the ``Reynolds number based on resistivity". } 
\begin{tabular}{ c c c c c c c c c} \hline  \hline
$\mathrm{Re}$ &  $N $ & $ N' $ & $\mathrm{Re_\sigma}$  &  $a$ & $D_N/D_\nu$ \\ \hline
10000 	& 0 	& 0 		  	& -  		& 5.0/3  	& 0 \\
10000 	& 1 	& 0.0028	 	& 1 		& 1.8  	& 1.8 \\
10000 	& 5 	& 0.014	 	& 0.2 	& 2.1  	& 15.0 \\ \hline

2500 	& 0 	& 0 		  	& -  		& 5.0/3  	& 0 \\
2500 	& 1 	& 0.0057		& 1		& 1.9  	& 1.35 \\
2500 	& 5 	& 0.028		& 0.2		& 2.2  	& 9.7 \\ \hline \hline
\end{tabular}

\label{Table1}
\end{table}

\section{Conclusions and Discussions}
In this paper we construct a model based on variable energy flux for modeling liquid-metal MHD under quasi-static approximation.  The energy flux decreases with the increase of wavenumbers due to the dissipation induced by the Lorentz force.  Since this dissipation is effective at all scales, the energy spectrum in the inertial range gets steepened significantly.  The magnitude of the spectral exponent increases monotonically with the interaction parameter $N$.  Our spectral exponents are in qualitative agreement with the numerical~\cite{Vorobev:2005PF,Burattini:2008PF,Burattini:2008PD} and experimental results~\cite{Kolesnikov:1974,Alemany:1979JM}.

Based on experiments and numerical simulations, some researchers argue that the energy spectrum of the quasi-static MHD is proportional to $k^{-3}$, similar to the forward enstrophy cascade regime of two-dimensional fluid turbulence.   However,  Favier {\it et al.}~\cite{Favier:2010PFb} showed that the velocity profile of quasi-static MHD turbulence differs significantly from 2D fluid turbulence, and it resembles ``two-and-half-dimensional-", or ``two-dimension three-component (2D-3C)" turbulence with all the three components of the velocity field being important.  We remark however that even for the aforementioned 2D-3C turbulence, the spectral exponent could be greater than 3 due to the additional dissipation induced by the Lorentz force, similar to the steepening of the energy spectrum for the two-dimensional turbulence with Ekman friction~\cite{Boffetta:2007EPL,Verma:2012EPL}.   Note that for fluid turbulence, the energy is conserved in the nonlinear triad interactions, and the dissipation takes place at small scales only.  Thus, the quasi-static MHD turbulence in which dissipation takes at all scales is very different from pure fluid turbulence.  A simple extrapolation of pure fluid turbulence to quasi-static MHD turbulence is not appropriate.    

The system under investigation is anisotropic due to the $\cos^2(\theta)$ term in the Lorentz force term.  We believe that the $g(\theta)$ term of Eq.~(\ref{eq:Ek}) could capture the anisotropic effects in a significant manner.   Unfortunately our present calculation assumes isotropic energy spectrum ($g(\theta) =1$, or $c_1=1$ and $c_2=1/2$) due to an inadequate knowledge of $g(\theta)$.   We are in the process of computing $g(\theta)$ using direct numerical simulations, which will enable us to compute precise values of $c_{1,2}$ for various values of the interaction parameters.  We expect the spectral exponents computed by this modified model to agree better with the numerical simulations and experiments.  For time being, the isotropic energy spectrum is expected to be applicable for moderate $N$'s ($N\sim 1$). We need to extend the present analysis to anisotropic situations, especially for large interaction parameters.

%Our spectral exponents are in qualitative agreement with the numerical~\cite{Vorobev:2005PF,Burattini:2008PF,Burattini:2008PD} and experimental results~\cite{Kolesnikov:1974,Alemany:1979JM}.  However, our exponents are somewhat lower than the corresponding exponents reported in experiments and numerical simulations.  These discrepancies could be due to inadequate modeling of the parameters $c_1$ and $c_2$ due to poor knowledge of $g(\theta)$.     Future numerical simulations would help us resolve these issues in a significant manner.  

\acknowledgments
I thank Daniele Carati, Bernard Knapen, Raghwendra Kumar, Sandeep Reddy, and P. Satyamurthy for useful comments and suggestions.  I am grateful to Mani Chandra for valuable help in python programming and plotting.  This work was supported by a research grant from Department of Science and Technology, India as Swarnajayanti fellowship, and a research grant 2009/36/81-BRNS from Bhabha Atomic Research Center.

%\bibliographystyle{phpf}
%\bibliography{/Users/mkv/res/bib/journal,/Users/mkv/res/bib/book}

\pagebreak
\newpage

\end{document}